\documentclass[twocolumn,prl,aps,floats,showpacs,superscriptaddress]{revtex4-1}
\usepackage{amsmath}

\usepackage{color}
\usepackage{epsfig}
\usepackage{textcomp}
\usepackage{gensymb}
\usepackage{graphicx}
\usepackage{dcolumn}
\usepackage{bm}
\usepackage{amssymb}

\def\gapp{\lower.35em\hbox{$\stackrel{\textstyle>}{\sim}$}}
\def\lapp{\lower.35em\hbox{$\stackrel{\textstyle<}{\sim}$}}

\begin{document}
\bibliographystyle{apsrev4-1}

\title{Emergence of a ZO Kohn anomaly in quasi-freestanding epitaxial graphene}
\author{Antonio Politano}
\affiliation{Dipartimento di Fisica, Universit\`a degli Studi della Calabria, 87036 Rende
(Cs), Italy}
\author{Fernando de Juan}
\affiliation{Materials Science Division, Lawrence Berkeley National Laboratories, Berkeley, CA
94720, USA}
\affiliation{Department of Physics, University of California, Berkeley, CA 94720, USA}
\author{Gennaro Chiarello}
\affiliation{Dipartimento di Fisica, Universit\`a degli Studi della Calabria, 87036 Rende
(Cs), Italy}
\affiliation{Consorzio Nazionale Interuniversitario di Scienze Fisiche della Materia, via della
Vasca Navale 84, 00146, Rome, Italy}
\author{Herbert A. Fertig}
\affiliation{Department of Physics, Indiana University, Bloomington, IN 47405, USA}
\date{\today}

\begin{abstract}
In neutral graphene, two prominent cusps known as Kohn anomalies are found in the phonon dispersion of the highest optical phonon at $q=\Gamma$ (LO branch) and $q=K$ (TO branch), reflecting a significant electron-phonon coupling (EPC) to undoped Dirac electrons. In this work, high-resolution electron energy loss spectroscopy is used to measure the phonon dispersion around the $\Gamma$ point in quasi-freestanding graphene epitaxially grown on Pt(111). The Kohn anomaly for the LO phonon is observed at finite momentum $q\sim2k_F$ from $\Gamma$, with a shape in excellent agreement with the theory and consistent with known values of the EPC and the Fermi level. More strikingly, we also observe a Kohn anomaly at the same momentum for the out-of-plane optical phonon (ZO) branch. This observation is the first direct evidence of the coupling of the ZO mode with Dirac electrons, which is forbidden for freestanding graphene but becomes allowed in the presence of a substrate. Moreover, we estimate the EPC to be even greater than that of the LO mode, making graphene on Pt(111) an optimal system to explore the effects of this new coupling in the electronic properties. 
\end{abstract}
\maketitle

\emph{Introduction} - Kohn anomalies are kinks in the phonon dispersion of a material produced by abrupt variations in the screening of atomic vibrations by gapless electrons~\cite{K59,BRZ09,AKB08}. The properties of these anomalies are completely determined by the electron-phonon coupling (EPC) and the shape of the Fermi surface, and their observation thus offers a window to study the interplay between electronic properties and phonon dynamics. 

As a paradigmatic example, in graphite~\cite{MRT04,PLM04,LPM06,MMD07} and graphene~\cite{LM06,ZSF08,THD08,SYM09,MMD09,MKD10,JF12a,JF12}, Kohn anomalies are realized as linear cusps in the dispersion of the highest optical phonon branches at $\Gamma$ (the $E_{2g}$ phonon) and at $K$ (the $A_1'$ phonon). The form and position of this cusp is determined by the Dirac fermion dispersion of the electronic $\pi$ bands around $K$ and the Fermi level $\mu \approx 0$. These two in-plane phonon branches are commonly accepted to be the only ones with a significant EPC in graphene~\cite{BPF09}. Within a tight-binding picture, these are the only branches that modify the nearest neighbor hopping integrals, so other in-plane phonons require a different mechanism to couple to electrons. 

For the case of out-of-plane phonons, the EPC is strongly constrained by the presence of mirror symmetry with respect to the horizontal plane which forbids a first order coupling to electrons, at least in free-standing graphene. When graphene is supported by a substrate a first order EPC becomes allowed in principle~\cite{L11,B08}, but it has remained unknown whether its strength is large enough to produce any significant effect. In particular, for the out-of-plane optical mode (the ZO mode), it has been speculated that a finite coupling could be responsible for a Peierls instability to a spontaneous buckling of the lattice~\cite{FL07,ZKM11}, which has not been observed so far. A coupling of the ZO mode to electrons would influence many physical properties, since the ZO has significantly lower energy than the in-plane branches with strong EPC. The observation of a Kohn anomaly for the ZO phonon~\cite{GP09} would represent a definitive proof of the existence of the EPC for this phonon branch and would allow an estimate of it. Herein, we demonstrate the existence of precisely this Kohn anomaly by means of high-resolution electron energy loss spectroscopy~\cite{PCB13} (HREELS) measurements on monolayer graphene grown on Pt(111). In addition, we also present evidence of the expected Kohn anomaly for the $E_{2g}$ phonon branch.  

Graphene on Pt(111) is an ideal playground for investigating the possible emergence of a ZO Kohn anomaly, as it is characterized by the weakest graphene-substrate interaction~\cite{GPH11}. The graphene-Pt distance (3.31 $\rm{\AA}$) lies close to the $c$-axis spacing in graphite. Moreover, the electronic structure of graphene on Pt(111) resembles that of isolated graphene~\cite{SSS09}, with the Dirac fermion dispersion of $\pi$ bands preserved. Angle-resolved photoemission spectroscopy (ARPES) experiments~\cite{SSS09} do not show any significant hybridization of the graphene $\pi$ states with Pt $d$-states, which simply superpose in energy with minimal interaction between them. This is in contrast to the case of, for example, graphene on Ni(111), where the hybridization of the graphene $\pi$-states with the Ni $d$ bands has a very strong effect on the $\pi$ bands~\cite{AW10}. The graphene sheet grown on Pt(111) is $p$-doped with a shift by $0.30\pm 0.15$ eV of the Fermi level from the Dirac point. 

\begin{figure}[t]
\begin{center}
\includegraphics[width=8cm]{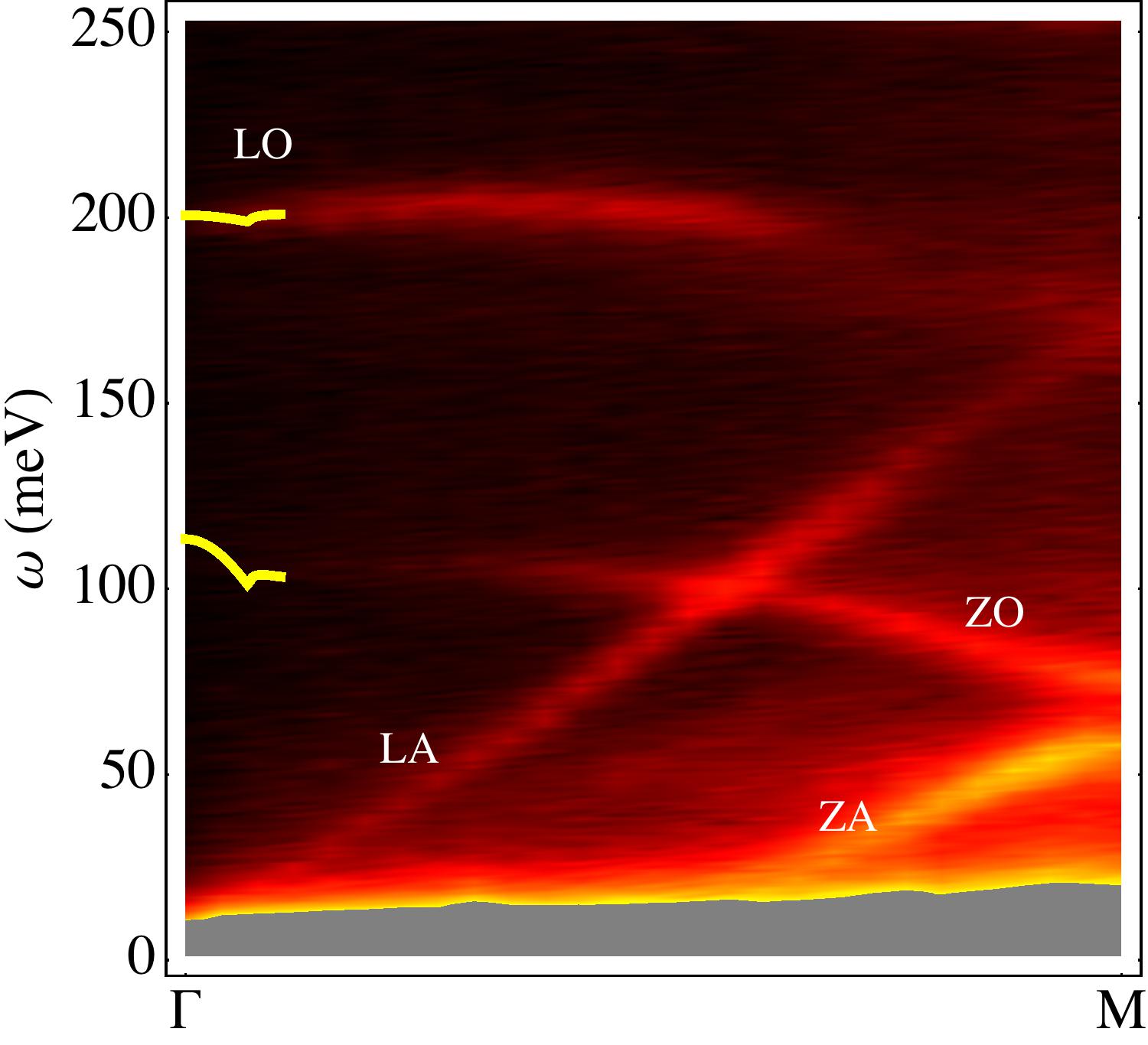}
\caption{HREELS intensity plot for phonon dispersion of graphene/Pt(111). Phonon modes have been recorded in HREELS spectra acquired in off- specular geometry with the sample oriented along the $\Gamma-M$ direction. The incidence angle is $\theta_i= 80 \degree$ with respect to the surface normal and the primary electron beam energy is $E_p = 20$ eV. To put in evidence only inelastic losses due to phonons, the tail of the elastic peak is grayed out for clarity. The LO, LA, ZO and ZA branches are clearly identified, while the TO and TA are barely visible due to a selection rule~\cite{JPC14}. Note the very low intensity of the ZO mode close to $\Gamma$. Yellow lines are fits to the dispersion obtained with higher resolution data, where Kohn anomalies can be appreciated (see text for details).}\label{EELSmap}.  
\end{center}
\end{figure}

\emph{Experimental methods} - Experiments were carried out in an ultra-high vacuum chamber operating at a base pressure of $5\cdot 10^{-9}$ Pa. The sample was a single crystal of Pt(111). The substrate was cleaned by repeated cycles of ion sputtering and annealing at 1300 K. Surface cleanliness and order were checked using Auger electron spectroscopy and low-energy electron diffraction measurements, respectively. Graphene was attained by dosing ethylene onto the clean Pt(111) substrate held at 1150 K. The completion of the first layer was reached upon an exposure of $3\cdot10^{-8}$ mbar for ten minutes (24 $L$, 1 $L=1.33\cdot10^{-6}$ mbar·s). The graphene layer was characterized by Raman spectroscopy, low-energy electron diffraction, and scanning electron microscopy experiments~\cite{CCC13,SM}. In particular, Raman measurements indicate the unique presence of monolayer graphene domains.

HREELS experiments were performed by using an electron energy loss spectrometer (Delta 0.5, SPECS) with an energy resolution ranging from 1 to 3 meV. The dispersion of the peaks in the energy loss $E_{\rm{loss}}$ was measured by moving the analyzer while keeping the sample and the monochromator fixed. The phonon in-plane momentum was determined from $\vec q_{\parallel} = \vec k_i\sin \theta_i-\vec k_s \sin \theta_s$ as
\begin{equation}
q_{\parallel} = \sqrt{2m E_p} \left(\sin \theta_i-\sqrt{1-E_{\rm{loss}}/E_p} \sin \theta_s \right)
\end{equation}
where $\theta_i$ and $\theta_s$ are the incident and scattering angles, and we set $\hbar=1$. The impinging energy $E_p$ and the incident angle $\theta_i$ were chosen so as to obtain the highest signal-to-noise ratio. A primary beam energy of $E_p$=20 eV provided the best compromise among surface sensitivity, the highest cross-section for mode excitation and momentum resolution. The angular acceptance of the apparatus was $\alpha=\pm0.5 \degree$, which determines the momentum resolution as
\begin{equation}
\Delta q_{\parallel} = \sqrt{2m E_p}\left(\cos \theta_i+\sqrt{1-E_{\rm{loss}}/E_p} \cos \theta_s \right)\alpha
\end{equation}
For the investigated range of $q_{\parallel}$, $\Delta q_{\parallel}$ was found to range from 0.005 near $\Gamma$ to 0.022 $\AA^{-1}$ at $K$. To obtain the energies of loss peaks, a polynomial background was subtracted from each spectrum. The resulting spectra were fitted by a Voigt line shape~\cite{SM}. All measurements were made at room temperature.
\begin{figure}[b]
\begin{center}
\includegraphics[width=8cm]{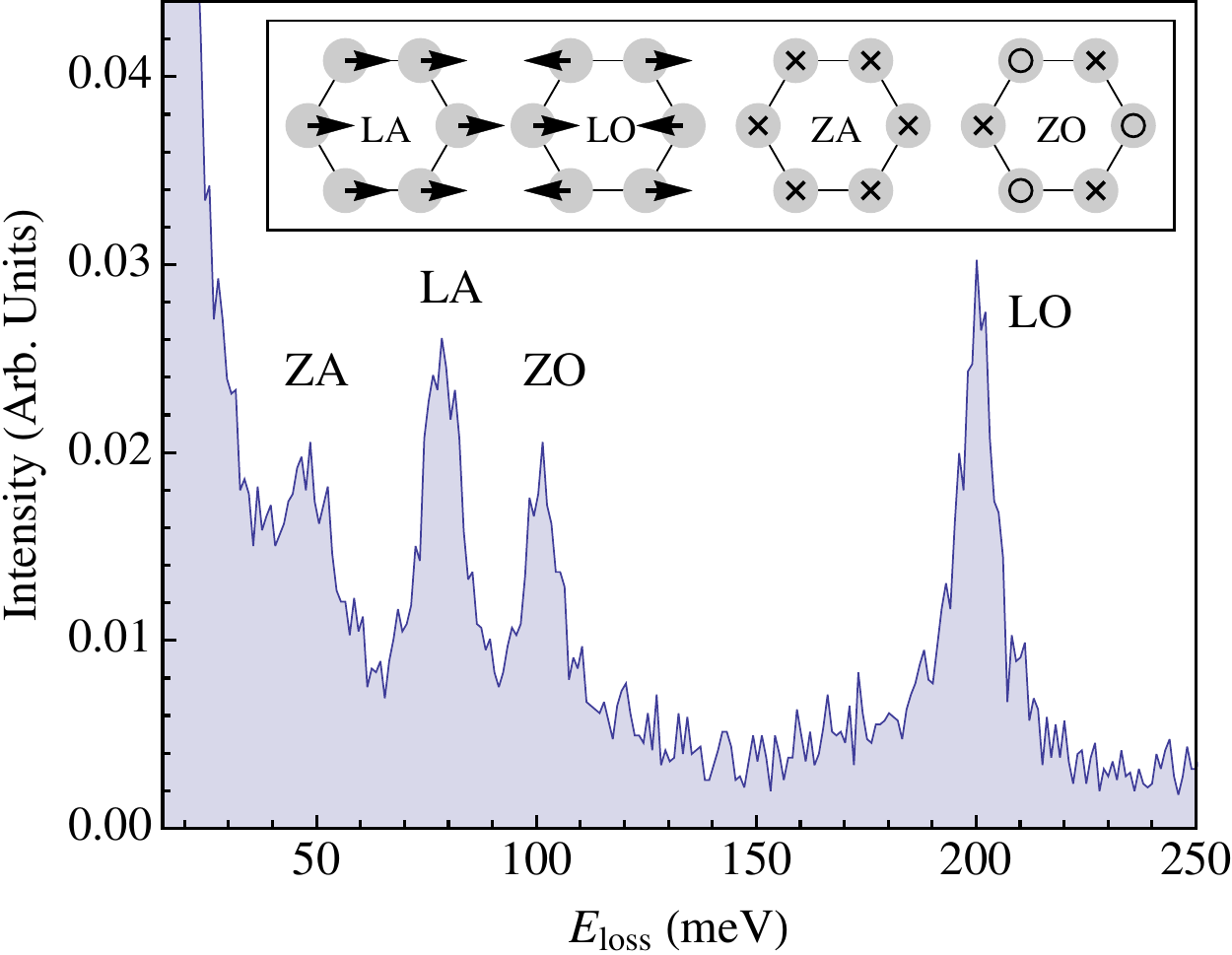}
\caption{Selected phonon spectrum from the dataset of Fig~\ref{EELSmap} at $\theta_s= 48 \degree$, which corresponds to $q_\parallel = 0.56$ ${\rm \AA}^{-1}$. The TA and TO modes are not observed due to a selection rule. Inset: Phonon displacements for the four phonons observed. Crosses and circles denote displacements in and out of the plane respectively.}\label{losspeaks}
\end{center}
\end{figure}

\begin{figure*}[t]
\begin{center}
\includegraphics[width=18cm]{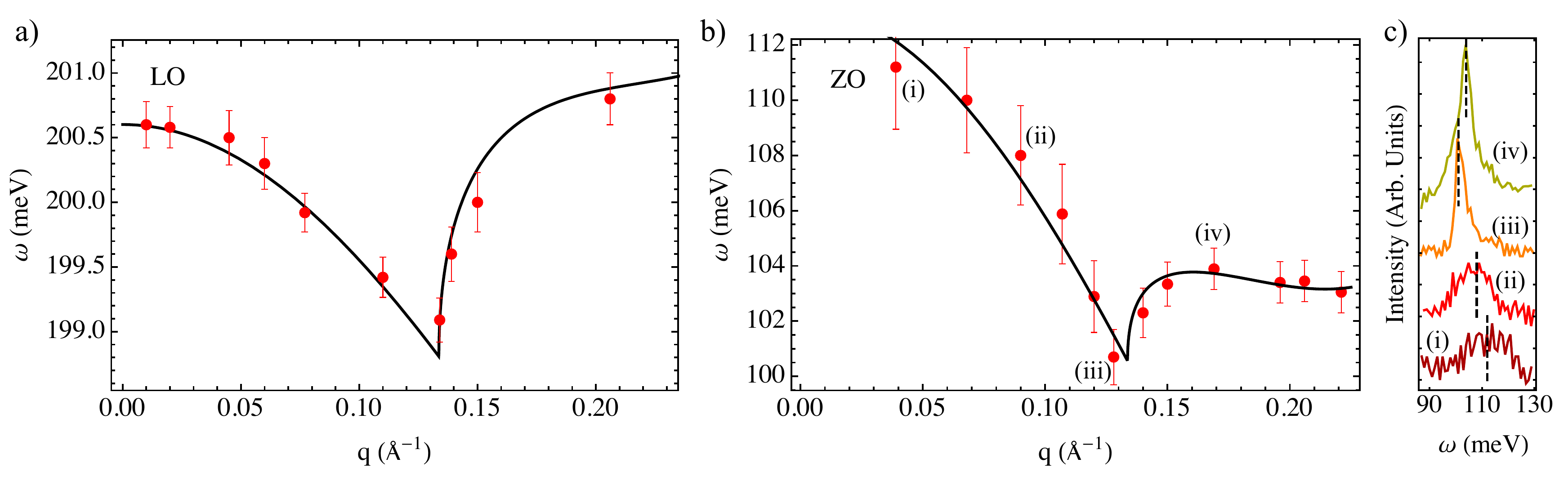}
\caption{Phonon dispersions close to the $\Gamma$ point, showing cusps around around $q_c\sim0.13$ $\AA^{-1}$. a) Dispersion of the LO phonon. A fit to Eq.~\ref{dispersion} (full line) gives an EPC $\lambda_{LO} = 0.029$, and parameters $\omega_{LO}^0 = 200.6$ meV, $a_{LO} =-110$ meV ${\rm\AA}^2$, $b_{LO} = 558$ meV ${\rm\AA}^4$. b) Dispersion of the ZO phonon. A fit to Eq.~\ref{dispersion} (full line) gives an EPC $\lambda_{ZO}=0.087$ and $\omega_{ZO}^0 = 102.3$ meV, $a_{ZO} =-810$ meV ${\rm\AA}^2$, $b_{ZO} = 5243$ meV ${\rm\AA}^4$. c) Representative spectra used to produce b). The frequencies are marked with a dashed line. Note the fits in a) and b) are also shown in Fig. \ref{EELSmap} as yellow lines.}\label{fits} 
\end{center}
\end{figure*}

\emph{Experimental results} - The phonon spectrum in graphene is composed by six phonon branches. Four vibrate in-plane and are labeled transverse and longitudinal acoustic (TA and LA) and optical (TO and LO). The other two are acoustic and optical out-of-plane vibrations (ZA and ZO). Fig.~\ref{EELSmap} shows a map of the HREELS intensity as a function of energy and momentum, where four phonon branches (ZA,ZO,LA,LO) can be clearly observed. Their dispersion is consistent with previous works~\cite{AW10,PMC12,EMW13}. The other two phonon branches (TO and TA) are practically absent because a selection rule that forbids the emission of odd phonons under reflections by the scattering plane~\cite{JPC14}. This is also clearly observed in Fig.~\ref{losspeaks}, where the energy loss is plotted for a selected angle, and four peaks are clearly identified. The displacement pattern for each of these four modes is shown as an inset to Fig.~\ref{losspeaks}.

We now focus on the dispersion of the ZO and LO modes around $\Gamma$. As usual, the ZO mode is significantly softened compared to the LO mode, due to the higher freedom for atom motion perpendicular to the plane with respect to the in-plane motion. The intensity of the ZO phonon at small momenta has very low weight, as shown in the intensity plot reported in Fig.~\ref{EELSmap} (see also~\cite{SM}). To enhance the signal-to-noise ratio from the ZO phonon in the nearness of $\Gamma$, each spectrum was acquired for several days. While this allowed us to resolve the ZO loss peak, the error bars for its frequency are still in general larger than for other branches. A careful fitting procedure~\cite{SM} was used to extract the experimental frequencies for the LO and ZO phonons for small momenta, which are reported in Fig.~\ref{fits}. The most striking feature in these plots is that they both display a clear cusp \emph{at the same momentum} $q \sim 0.13 \rm{\AA}^{-1}$. This strongly suggests that both cusps are Kohn anomalies at $q=2k_F$, which originate from the interaction of phonons with electrons. The Fermi wave-vector $k_F = E_F/v_F$ can be estimated from ARPES measurements of graphene/Pt(111)~\cite{SSS09}. The reported values of the Fermi energy and Fermi velocity are $E_F \approx 0.30 \pm 0.15$ eV and $v_F \approx 6$ eV$\rm{\AA}$. Thus, $2k_F \approx 0.10 \pm 0.05  \rm{\AA}^{-1}$, in good agreement with the position of the cusp found in this work. The position of the G peak in our Raman measurements~\cite{SM} can be used to estimate~\cite{F07,DPC08} $E_F \sim 0.42$ eV, also in good agreement. While a Kohn anomaly for the LO mode is known to occur, the presence of the cusp for the ZO mode is unexpected and represents the first evidence of the coupling of the ZO mode to electrons. In hindsight, by carefully inspecting the phonon dispersion recorded for graphene on Pt(111) in a previous experiment by some of the authors~\cite{PMC12}, a dip for both LO and ZO at finite momentum is in fact seen but was not noticed or discussed. In Ref.~\cite{PMF11}, the LO anomaly was incorrectly presented by shifting it to the $\Gamma$ point due to a misunderstanding, but in fact occurs at finite momentum as well. To substantiate the claim that these cusps are in fact Kohn anomalies, we now compute the self-energies of both LO and ZO phonons following the conventions of Refs.~\cite{BA08,BPF09,JF12a} and compare the predictions to the experimental data. 

\emph{Theory} - The dispersion of the phonons around $\Gamma$ is modified because of their coupling to electrons, which at low energies can be modeled with a Dirac Hamiltonian
\begin{equation}\label{H_el}
H = \int \frac{d^2k}{(2\pi)^2} \psi^{\dagger} \left( v_F \vec \sigma \cdot \vec k -\mu \right)\psi,
\end{equation}
with $\vec \sigma = (\sigma_x, \sigma_y)$ are Pauli matrices, $v_F$ is the Fermi velocity and $\mu$ is the chemical potential. This effective model is applicable up to energies $\Lambda_E \approx$ 1.5 eV or momenta $\Lambda_q = \Lambda_E / v_F =$ 0.25 $\AA ^{-1}$, beyond which the dispersion is no longer linear. Therefore, our predictions are only valid for phonon momenta within this range as well. The phonon Hamiltonian is
\begin{align}
H = \sum_i \int \frac{d^2q}{(2\pi)^2} \omega_{i}(q) b^{\dagger}_{i,q} b_{i,q},\label{phononH}
\end{align}
with creation and destruction operators defined by the effective displacements associated with each phonon mode
\begin{align}
u_i =  \sqrt{\frac{A_c}{4 \omega_i M}} \int \frac{d^2q}{(2\pi)^2} (b_{i,q}e^{i \vec q \vec r} +
b^{\dagger}_{i,q}e^{-i \vec q \vec r}),\label{phononDisp}
\end{align}
where $i=LO,TO,ZO$ and $A_c = 3\sqrt{3}a^2/2$ is the unit cell area, with $a=1.42 \rm{\AA}$ the nearest neighbor distance. The dispersions of the phonons $\omega_i(q)$ are analytic in the absence of EPC. For momenta close to the $\Gamma$ point they can be expanded as $\omega_i(q) = \omega^0_i + a_i q^2 +b_i q^4$, where $\omega_i^0$, $a_i$ and $b_i$ are parameters to be fitted from experimental data.  The coupling between electrons and phonons is described by
\begin{equation}
H_{e-ph} =  F_i \int d^2 r u_i \psi^{\dagger} \mathcal{M}_i \psi,\label{epv}
\end{equation}
with $F_i$ the electron-phonon coupling. The matrix $\mathcal{M}_i$ for the different phonons is equal to $\mathcal{M}_{LO}= \hat q \times \vec \sigma $, $\mathcal{M}_{TO}=\hat q \cdot \vec \sigma$, $\mathcal{M}_{ZO}=\sigma_3$. Following Ref.~\cite{BPF09}, it is customary to introduce a dimensionless EPC as 
\begin{equation}
\lambda_i =\frac{F_i^2 A_c}{2 M \omega_i v_F^2}\label{dimensionless}
\end{equation}
The electron-phonon coupling induces a phonon self-energy $\Pi_i(q)$ that corrects the dispersion according to
\begin{equation}
\omega_{R,i} = \omega^0_i + a_i q^2 +b_i q^4 +\frac{\lambda_i}{2}\Pi_i(q/k_F).\label{dispersion}
\end{equation}
In the static approximation, the self-energies for the different phonon branches in terms of the dimensionless variable $x=q/k_F$ are given by~\cite{SM}
\begin{align}
\Pi_{LO}(x) &= 
\frac{g_{s,v} \mu}{4\pi} \left(\sqrt{1-\tfrac{4}{x^2}} +\tfrac{x}{2} \arccos(2/x)\right)\theta(2-x),
\\
\Pi_{TO}(x) &= 0,\\
\Pi_{ZO}(x)& = 
\frac{g_{s,v} \mu}{4\pi}  \left(2 +x \arccos(2/x)\theta(2-x)\right).
\end{align}
These expressions can be used to fit the cusps in the experimental curves to obtain an estimate for the different electron-phonon couplings. To do so, we have fitted the dispersion parameters for both curves and determined the optimal Fermi level as $E_F = v_F q_c/2 \approx 0.401$ eV, which corresponds to a cusp momentum of $q_c=0.133$ $\AA^{-1}$. The fits are shown in Fig.~\ref{fits} with the experimental dispersion, and the result for the LO EPC is $\lambda_{LO} = 0.029$, in excellent agreement with the estimates obtained from Raman, $\lambda_{LO} = 0.027-0.034$~\cite{BPF09}. The fit for the ZO gives $\lambda_{ZO}=0.087$, an even greater value.

\emph{Discussion} - Since this is the first work to observe the effects of electron-phonon coupling to the ZO mode in graphene, there are no measurements of $\lambda_{ZO}$ available to compare with. Theoretically, this coupling has only been estimated for a SiO$_2$ substrate in Ref.~\cite{FL07}, giving a maximum value of $F_{ZO} = 7$ $\rm{eV \AA}$ or $\lambda_{ZO} =0.011$ \footnote{In the notation of Ref.~\cite{FL07}, when the sublattice displacements are $h_i=(\eta,-\eta)$, the mass at the Dirac point is $M = D \eta$. In our case $M = F_{ZO} u_{ZO}$, and the displacement is $h_i=u_{ZO}(1,-1)/\sqrt{2}$ because we use normalized phonon eigenstates. This implies $F_{ZO} = D/\sqrt{2}$. With their estimate of $Da = 1-14$ eV and $a=1.42$ $\rm{\AA}$ we obtain $F_{ZO} = 0.5-7$ $\rm{eV \AA}$}. Our significantly higher value reflects the fact that the substrate is metallic and that, while direct hybridization with graphene is small, the surface electric field induces a stronger breaking of reflection symmetry which is responsible for the EPC.

Our finding that $\lambda_{ZO}$ has a significant magnitude implies that its effects should also be observable in other experiments. For example, the EPC is visible in the electron dispersion, in the form of a kink at the phonon frequency that is observable by ARPES~\cite{ZSF08}. The existence of $\lambda_{ZO}$ would be responsible for an extra kink at $\omega_{ZO}$ that could be resolved in future experiments. One could also expect a finite $\lambda_{ZO}$ to induce new Raman peaks. A peak at $\omega_{ZO}$ from a first order Raman process (an analog of the $G$ peak) for the ZO is not allowed because the ZO transforms as $B_2$, which is not contained in $E_1 \times E_1$. However, a second order process (analogous to the $2D'$ peak~\cite{BPF09}) is in principle allowed by symmetry. This process would give a peak at roughly $2\omega_{ZO}$, which unfortunately may be difficult to detect because of its overlap with the $G$ peak. Therefore, either Raman or ARPES could provide an independent confirmation of electron-phonon coupling for the ZO mode. 

The finding that the substrate can induce an EPC to the ZO may also have important implications for transport in the high-field regime. The phenomenon of current saturation at high bias originates from inelastic scattering with optical phonons, both intrinsic~\cite{BLM09} or from the substrate~\cite{MHY08}. Since the energy of the ZO ($\sim 100$ meV) is significantly lower than the known phonons with strong EPC, the effects of the ZO may already be present in current experiments~\cite{MHY08,BLM09,PA10,DZJ10}, and its presence should be accounted for in theory~\cite{BM09,TS09,FKA11}.  

A related system where the standard (LO and TO) Kohn anomalies have been observed is graphene on Ir(111)~\cite{EMW13}. According to ARPES~\cite{PKP09}, the Fermi level in this system is $E_F\approx 0.1$ eV which corresponds to $2k_F\approx0.03$ $\rm{\AA}^{-1}$. Kohn anomalies at such small momentum are probably too difficult to resolve, which is consistent with the observed broadened dip around $\Gamma$ instead of a cusp. Similarly, if the ZO EPC in this system is significant, such a feature should be observed for the ZO around $\Gamma$. The data in Ref.~\cite{EMW13} are inconclusive on this matter. Finite momentum Kohn anomalies will be more likely to be found in graphene on metals more similar to Pt(111), with weak hybridization but large charge transfer~\cite{KGR09}.
In metallic substrates with stronger hybridization such as as Ni(111), Kohn anomalies are not present for the LO and TO phonons~\cite{SFA99,FRS00,FSR99,SFR98,ASI90} because the coupling of the graphene $\pi$ bands with the Ni $d$ orbitals completely rearranges the electron bands. A Kohn anomaly for the ZO is therefore not expected in this type of system. 

In summary, in this work we have shown that the presence of a substrate induces a significant electron-phonon coupling to ZO phonons in graphene, which is responsible for a strong Kohn anomaly at $q=2k_F$. This finding paves the way to explore the many implications of the coupling of flexural phonons to Dirac electrons in supported graphene samples and graphene-metal contacts. 

\emph{Acknowledgments}
We thank Davide Campi and Sinisa Coh for helpful discussions. A.P. and G.C. thank Fabio Vito for technical support. F. de J. acknowledges support from the "Programa Nacional de Movilidad de Recursos Humanos" (Spanish MECD). This work was supported in part by the US-Israel Binational Science Foundation.

\bibliography{kohn}

\pagebreak

\onecolumngrid
\vspace{0.2in}
\begin{center}
{\bf \large Supplementary Material: Emergence of a ZO Kohn anomaly in quasi-freestanding epitaxial graphene}
\end{center}
\vspace{0.1in}

\renewcommand{\thetable}{S\Roman{table}}
\renewcommand{\thefigure}{S\arabic{figure}}
\renewcommand{\thesubsection}{S\arabic{subsection}}
\renewcommand{\theequation}{S\arabic{equation}}

\setcounter{secnumdepth}{1}
\setcounter{equation}{0}
\setcounter{figure}{0}
\setcounter{section}{0}

\section{Review of experimental techniques}

In this section we compare the different experimental techniques to measure phonons and explain why the ZO Kohn anomaly can only be observed with HREELS. There are four basic techniques to probe phonons: Neutron Inelastic Scattering, Raman scattering of X-Rays, Inelastic He Atom Scattering and Electron Energy Loss Spectroscopy (EELS).

The first two techniques are better suited to the study of bulk 3D phonons. Neutrons are particularly well suited for bulk phonon dispersion studies because of their very small cross section with matter which allows them to penetrate deeply into the crystal. X-ray photons in grazing incidence, available at synchrotron radiation sources, are also a standard probe for bulk phonons and have in fact been used to study phonon dispersion in bulk graphite~\cite{MMD07}. But none of this methods is suitable for studying phonon modes of a single layer of graphite because of the lack of surface sensitivity. 

Inelastic helium atom scattering is a more appropriate surface probe. However, in standard conditions the impinging energy is 65 meV. In order to observe modes at 100 meV it would be necessary to keep the nozzle at very high temperature, which implies a strong decrease of the signal. Inelastic helium atom scattering is more appropriate to study phonons up to 20-30 meV. The ZO phonon at around 100 meV has too high energy to be probed by inelastic helium atom scattering.

HREELS measurements, on the other hand, have no limit in the energy loss and can easily access the high frequency range. They are surface sensitive measurements and have excellent resolution in both momentum and energy. This is thus the most suitable technique to study the full surface phonon spectrum, and in particular the only one capable of detecting the presence of the ZO Kohn anomaly. 

\begin{figure}[h]
\begin{center}
\includegraphics[width=16cm]{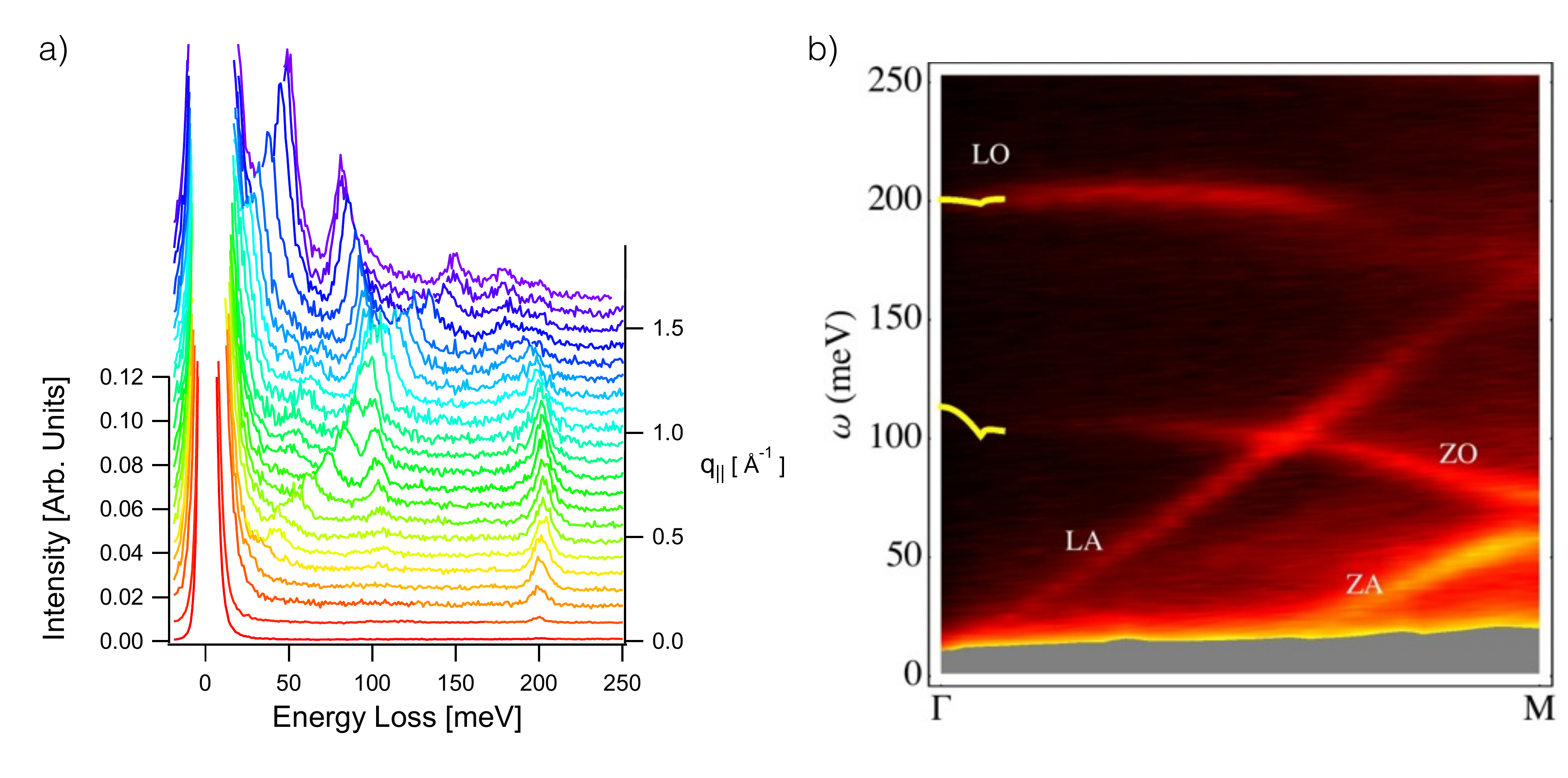}
\caption{a) Selected HREELS spectra for different values of the parallel momentum transfer $q_\parallel$, evaluated by inserting the kinematic conditions and the loss energies in Eq. 1 of the main text. The intensity of the various phonons has been normalized to that of the elastic peak. Note the poor intensity of the ZO mode near $\Gamma$. b) EELS Intensity plot for phonon dispersion of graphene/Pt(111), reproduced from Fig. 1 of the main text. Note again the very low intensity of the ZO mode close to $\Gamma$. The fits in Fig. 3 of the main text are displayed as yellow lines.}\label{sup1}
\end{center}
\end{figure}

\section{Fit procedures}

In this section we present more details of the process employed to obtain the phonon dispersions displayed in the main text, and in particular in Fig. 3 of the main text where the Kohn anomalies are demonstrated. First, the intensity plot in Fig. 1 of the main text was produced from many different spectra acquired at fixed incidence angle ($\theta_i = 80 \degree$, all angles with respect to the sample normal) and at scattering angles varying from $\theta_s = 80 \degree$ up to $15 \degree$. A representative subset of this series of spectra is shown in Fig. \ref{sup1} as a waterfall plot, along with the intensity plot which is reproduced here for ease of comparison. 

\begin{figure}[t]
\begin{center}
\includegraphics[width=9cm]{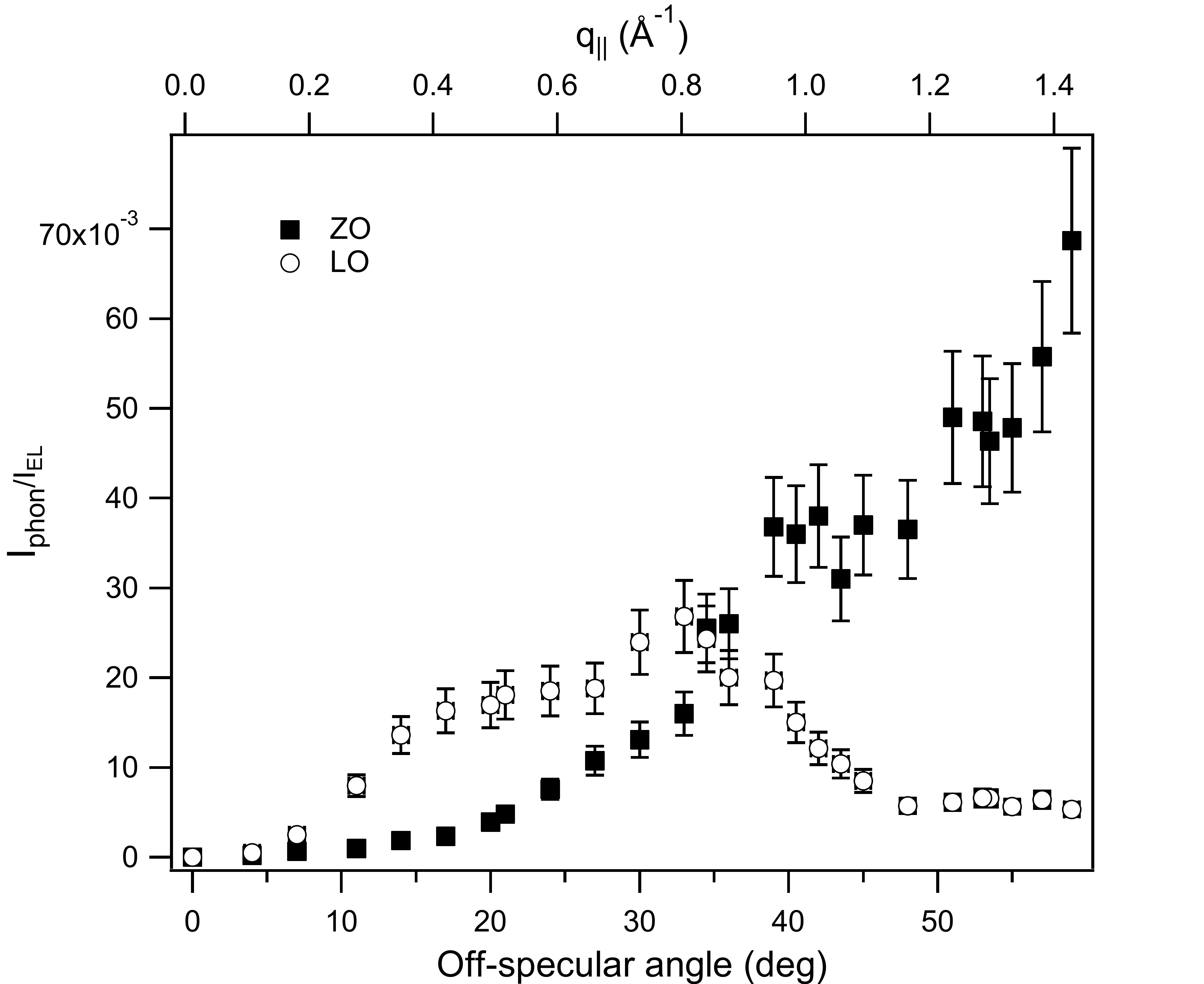}
\caption{Intensity of the ZO (filled squares) and the LO (empty circles) phonon modes as a function of the off-specular scattering angle, with the incidence angle fixed at $80 \degree$ with respect to the sample normal. The intensity of phonon modes has been normalized to that of the elastic peak.
}\label{intensity}
\end{center}
\end{figure}

It is clear that finding the location of the Kohn anomalies in an intensity plot by simple inspection is not really possible due to the low intensity near $\Gamma$. To help the reader locate the Kohn anomalies in the dispersion of the LO and ZO in the intensity plot, we have also included the fits of Fig. 3 of the main text as yellow lines in Fig. \ref{sup1}(b). To quantify the low intensity, in Fig.~\ref{intensity} we also present the integrated weight of the ZO and LO peaks as a function of the off-scattering angle with incidence angle kept at $80 \degree$. The intensity of the ZO mode increases with the off-specular angle, and thus with the momentum $q_\parallel$, with a continuous behavior. Instead, the LO mode show maximum intensity for $15 \degree$-$40 \degree$ off-specular angles. Near $\Gamma$, i.e., near the specular conditions $\theta_i=\theta_s = 80 \degree$, the intensity of the LO and especially of the ZO mode is extremely poor. The spectra near $\Gamma$ actually required several days of measurement in order to reach a sufficient signal-to-noise ratio. 

To quantify the dispersion of the LO and ZO around $\Gamma$ accurately, we extracted the energy for each peak in each spectrum with a given $q_\parallel$ with the following fitting procedure, exemplified in Fig.~\ref{fits}. First, an exponential background was subtracted from the raw HREELS spectrum. Then, the obtained spectrum was fitted with Voigt line-shapes. The Voigt line-shape is a convolution of Lorentzian and Gaussian line-shapes. The Lorentzian part takes into account the intrinsic line-shape of phonon modes, while the Gaussian line-shape is necessary to account for the experimental broadening due to the energy resolution of spectrometer. The extracted set of pairs $(E,q_\parallel)$ was then fitted with the model dispersion relations discussed in the text. Both the data and the fits are presented in Fig. 3 of the main text. 

\begin{figure}[t]
\begin{center}
\includegraphics[width=14cm]{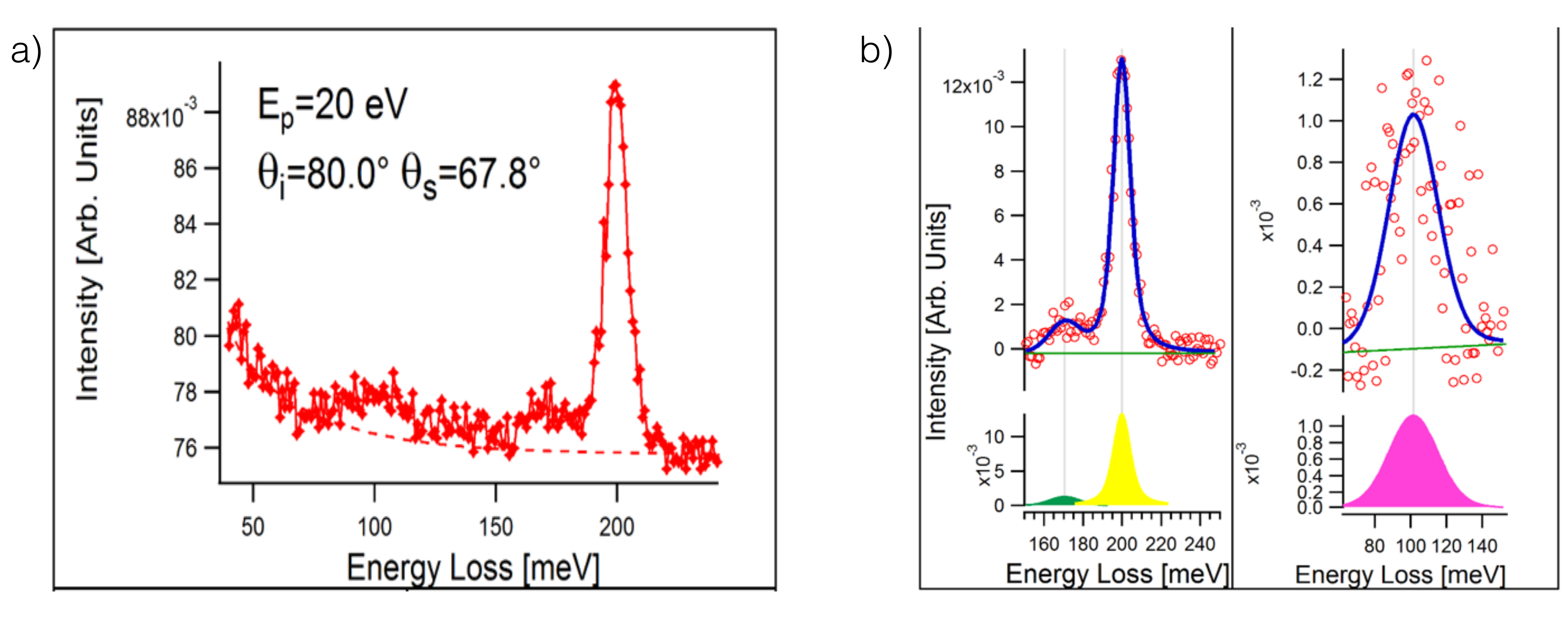}
\caption{a) Selected HREELS spectrum. The fit to an exponential background is reported with a dashed line. b) Fit procedures for TO+LO (left panel) and ZO modes (right panel) with the exponential background in a) substracted.}\label{fits}
\end{center}
\end{figure}

\section{Sample characterization}
\subsection{Scanning electron microscopy investigation}
In Fig. \ref{SEMRaman}(a) we report a scanning electron microscopy (SEM) of the graphene/Pt(111) sample, also reported elsewhere \cite{CCC13}. The surface morphology appears to be homogeneous across all the sample (1x1 cm$^2$). The SEM image shows a full coverage of the substrate surface by the graphene which forms a network of wrinkles (darker horizontal and vertical lines in Fig.~\ref{SEMRaman}(a)). The wrinkles network develops on a micrometric length scale. Its morphology matches that obtained by low-energy electron microscopy measurements for graphene grown on Pt(111) by carbon segregation from the Pt(111) substrate and other metallic substrates \cite{SSS09}.

\subsection{Raman investigation}
\begin{figure}[t]
\begin{center}
\includegraphics[width=17cm]{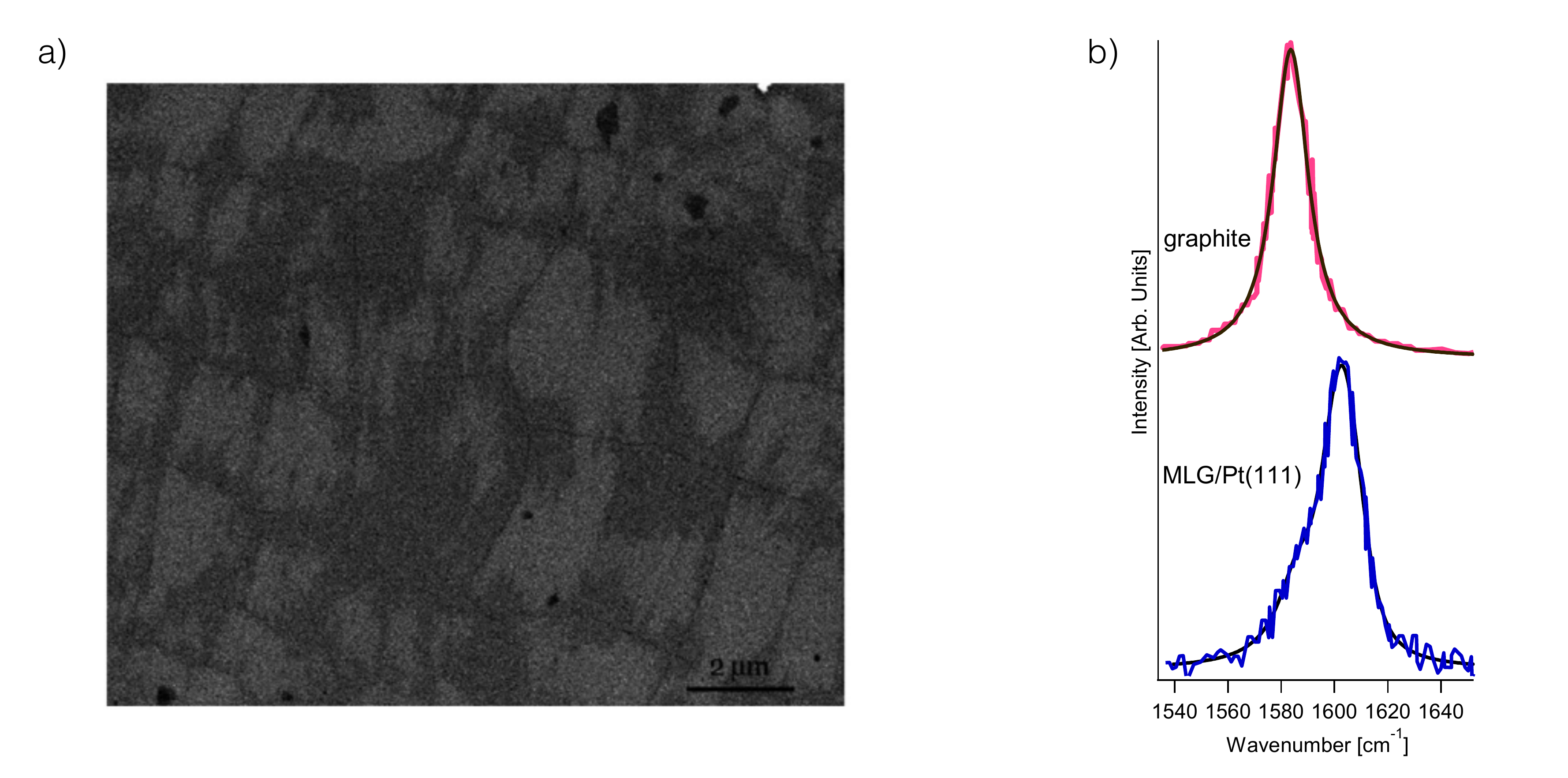}
\caption{a) SEM image of a region of the graphene layer on a Pt(111) substrate. Taken from Ref. \cite{CCC13}. b) Raman spectrum of graphite (data taken from Ref.~\cite{F07}) and MLG/Pt(111) (data taken from Ref.~\cite{CCC13})}\label{SEMRaman}
\end{center}
\end{figure}

The sample has been also analyzed ex-situ by Raman spectroscopy. Such experiments indicate that the surface of the Pt(111) substrate is well coated with monolayer graphene, which is more ordered and homogeneous than for similar preparations previously reported in literature. A compressive strain is found for the graphene as the effect of the growth process. For more details on Raman investigation, see Ref.~\cite{CCC13}.
In Fig.~\ref{SEMRaman}(b) we report a comparison between the Raman signal from the G peak in graphite and graphene/Pt(111). Graphite shows a single peak (data taken from Ref.~\cite{F07}), which can be fitted by a single Voigt line-shape. By contrast, graphene/Pt(111) shows a clearly asymmetric G peak, located at $\omega = 1603$ cm$^{-1}$. The position of the G peak can be used to estimate the Fermi level.  A peak at $1603$ cm$^{-1}$ corresponds to a hole concentration~\cite{DPC08} of $n = 1.5 \cdot 10^{13}$ cm$^{-2}$, which corresponds to $\mu = v_F (n\pi)^{1/2} \sim 0.42$ eV, in very good agreement with the estimate obtained from the position of the cusp, and within the error of the ARPES estimate $\mu = 0.3 \pm 0.15$ eV. 

It is interesting to note that the asymmetry of the Raman G peak could be originated in the emergence of a second order process with a peak around $2\omega_{ZO}$, activated by the linear electron-phonon coupling. However, since this type of second order peak can probe a range of energies because momentum is not restricted to $\vec k=0$, further theoretical modeling is required to determine unambiguously that this peak indeed originates from the ZO. 

\section{Theory: Phonon self-energy}

All calculations are performed at $T=0$. Room temperature (300 K) corresponds to 25 meV, which is much smaller than $E_F$ or the optical phonon frequencies. With the definitions given in Eqs.~\ref{phononH}-\ref{dimensionless} in the main text, the phonon propagator is written as
\begin{equation}\label{phonon}
G_{i}(\omega,q) = \frac{2\omega_i(q)}{\omega^2 - \omega_i(q)^2 -2\omega_i(q)\Sigma_i(\omega,q)},
\end{equation}
where $\Sigma_i(\omega,q) = \frac{\lambda_i}{2} \Pi_i(\omega,q)$ is the self-energy, and $i=LO,TO,ZO$ as in the text. The dispersion relation for the phonon can be obtained by solving for the pole in Eq. (\ref{phonon}) for small $\lambda_i$, so that the renormalized phonon dispersion relation is given by
\begin{equation}
\omega_{R,i}(q) \approx \omega_i(q) + \frac{\lambda_i}{2}\Pi_i(\omega_i,q). \label{omega1}
\end{equation}
which yields Eq.~\ref{dispersion} in the text. The response function $\Pi_i$ for a particular phonon is given by 
\begin{equation}
\Pi_i(E,\vec q) = g_{s,v}i\int \frac{d \omega d^2k}{(2\pi)^3} \;{\rm tr} \; \mathcal{M}_i G(\omega,\vec k)\mathcal{M}_i  G(\omega+E,\vec k+\vec q),\label{resp}
\end{equation}
where $g_{s,v}=4$ accounts for spin and valley degeneracy, $\mathcal{M}_i$ is the electron-phonon coupling matrix as defined in the text, and  
\begin{equation}
G(\omega,\vec k) = \frac{\omega+\mu +\vec \sigma \cdot \vec k}{(\omega(1+i \epsilon)+\mu)^2 -k^2},
\end{equation}
is the electron propagator, where we set $v_F=1$. Evaluating the trace in Eq.~\ref{resp} we obtain
\begin{equation}
\Pi_i(E,\vec q) = g_{s,v}i\int \frac{d \omega d^2k}{(2\pi)^3} \frac{N_i}{[(\omega(1+i \epsilon)+\mu)^2 -k^2][((\omega+E)(1+i \epsilon)+\mu)^2 -|\vec k+\vec q|^2]},
\end{equation}
where the numerators $N_i$ for the different phonons are given by
\begin{align}
N_{LO} =& 2[\omega(\omega+E) - \vec k \cdot (\vec k +\vec q) + 2(\vec k \cdot \vec q)^2/q^2 +2\vec k \cdot \vec q],\\
N_{TO} =& 2[\omega(\omega+E) - \vec k \cdot (\vec k +\vec q) + 2(\vec k \times \vec q)^2/q^2 ],\\
N_{ZO} =&  2[\omega(\omega+E) - \vec k \cdot (\vec k +\vec q)],
\end{align}
The frequency integral is performed with the residue method, giving 
\begin{equation}
\Pi_i(E,\vec q) = g_{s,v}\int \frac{d^2k}{(2\pi)^2} \left[\frac{ \theta(k-\mu)N_i |_{\omega = k-\mu}}{2|\vec k| [(k+E(1+i \epsilon))^2 -|\vec k+\vec q|^2]}+\frac{ \theta(|\vec k+\vec q|-\mu)N_i |_{\omega = -E + |\vec k+\vec q|-\mu}}{2|\vec k+\vec q| [(|\vec k+\vec q|+E(1+i \epsilon))^2 -|\vec k|^2]}\right],
\end{equation}
Since these integrals are ultraviolet divergent as power counting reveals, it is customary to split them into the contribution coming from the undoped Dirac cone plus a finite correction induced by finite chemical potential. This is done by replacing $\theta(k-\mu) = 1 - \theta(\mu-k)$ in the integrals, giving
\begin{equation}
\Pi_i(E,\vec q) = \Pi_i^{\mu=0}(E,\vec q) -\Delta \Pi_i(E,\vec q),\label{splitpi}
\end{equation}
with
\begin{equation}
\Delta \Pi_i(E,\vec q) =  g_{s,v}\int \frac{d^2k}{(2\pi)^2} \left[\frac{ \theta(\mu-k)N_i |_{\omega = k-\mu}}{2|\vec k| [(k+E(1+i \epsilon))^2 -|\vec k+\vec q|^2]}+\frac{ \theta(\mu-|\vec k+\vec q|)N_i |_{\omega = -E + |\vec k+\vec q|-\mu}}{2|\vec k+\vec q| [(|\vec k+\vec q|+E(1+i \epsilon))^2 -|\vec k|^2]}\right].
\end{equation}
$\Pi_i^{\mu=0}(E,\vec q)$ can then be computed with dimensional regularization methods giving~\cite{PLM04,BPF09}
\begin{align}
\Pi_{LO}^{\mu=0} =&   \frac{g_{s,v}}{16} \sqrt{-E^2+q^2}, \\
\Pi_{TO}^{\mu=0} =& -\frac{g_{s,v}}{16} \frac{E^2}{\sqrt{-E^2+q^2}},\\
\Pi_{ZO}^{\mu=0} =& \frac{g_{s,v}}{8}\sqrt{-E^2+q^2}.
\end{align}
The correction terms become complicated expressions but can be computed analytically in the static approximation ($E=0$)  giving \cite{THD08}
\begin{align}
\Delta\Pi_{LO} =& \frac{g_{s,v}}{4\pi} \left[\frac{q\pi}{4}-\left(\frac{q}{2} \arccos(\tfrac{2\mu}{q})+\frac{\mu}{q}\sqrt{q^2-4\mu^2}\right)\theta(q-2\mu)\right], \\
\Delta\Pi_{TO} =& 0, \\
\Delta\Pi_{ZO} =& \frac{g_{s,v}}{4\pi} \left[\frac{q\pi}{2}-2\mu-q \arccos(\tfrac{2\mu}{q})\theta(q-2\mu)\right] .
\end{align}
Combining the two contributions, Eq.~\ref{splitpi}, we finally get
\begin{align}
\Pi_{LO}(E=0) =& \frac{g_{s,v}}{4\pi} \left[\frac{q}{2} \arccos(\tfrac{2\mu}{q})+\frac{\mu}{q}\sqrt{q^2-4\mu^2}\right]\theta(q-2\mu), \label{fin1}\\
\Pi_{TO}(E=0)=& 0, \label{fin2} \\
\Pi_{ZO}(E=0) =& \frac{g_{s,v}}{4\pi} \left[2\mu+q \arccos(\tfrac{2\mu}{q})\theta(q-2\mu)\right]. \label{fin3}
\end{align}
The results in the main text are obtained by recovering $v_F$ and setting $x=q/(v_F\mu))$. Note that the functional dependence of the $\mu=0$ part for the ZO was anticipated in Ref. \cite{GP09}, and by symmetry arguments \cite{JF12}, it is the same as that of the TO phonon at the $K$ point \cite{B08}. 

\begin{figure}[t]
\begin{center}
\includegraphics[width=8cm]{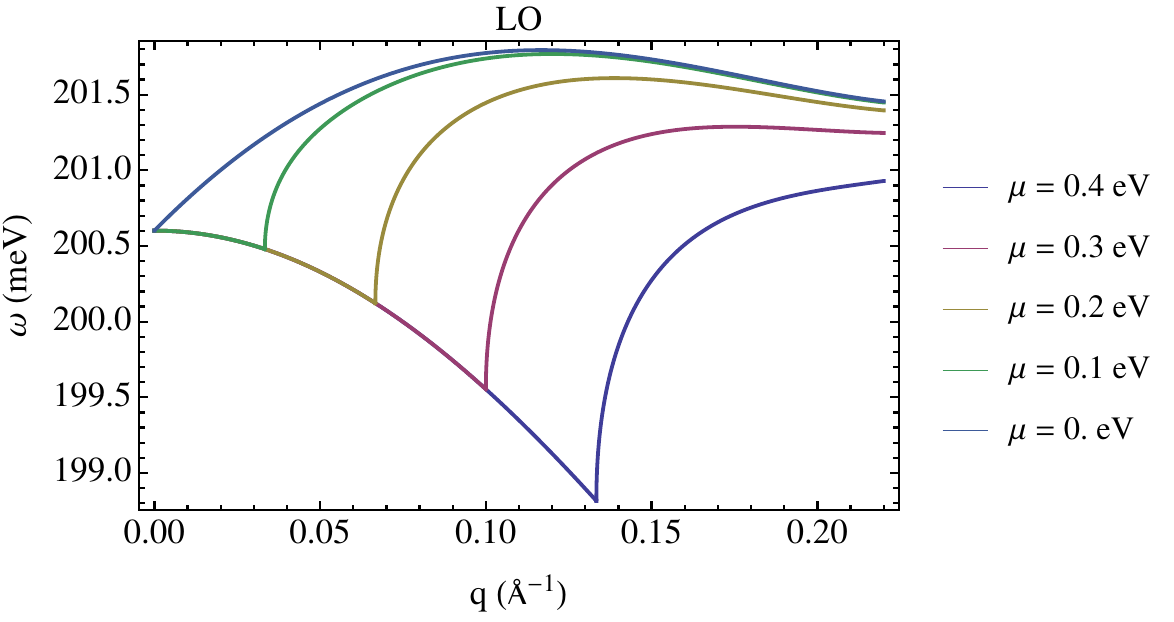}
\includegraphics[width=8cm]{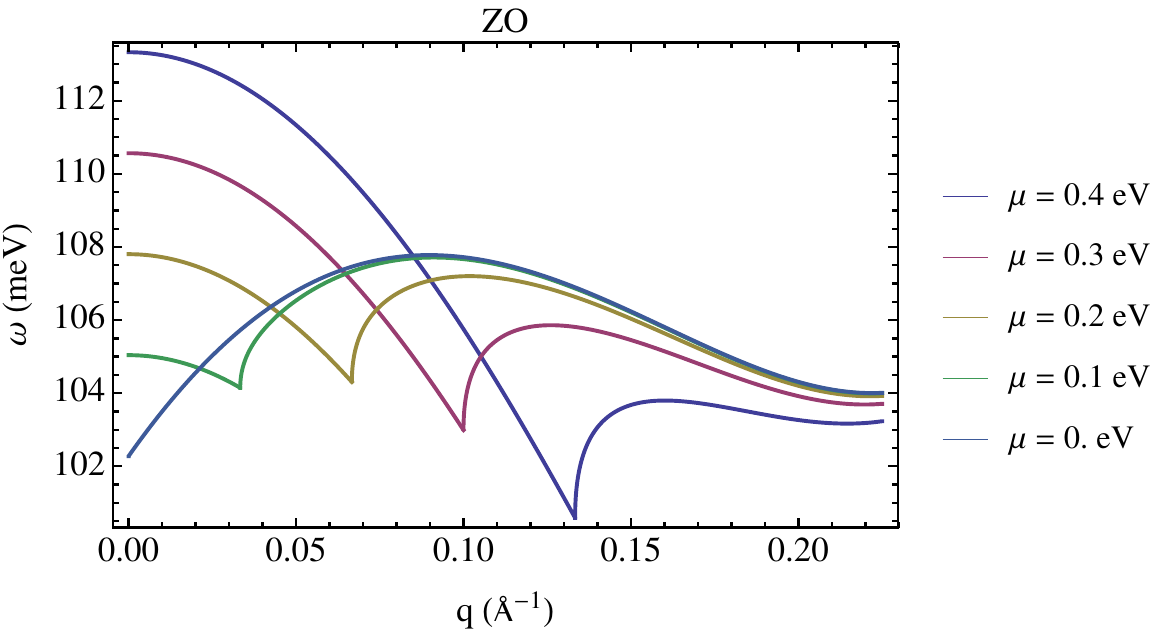}
\caption{Left: $\omega_{LO}(q)$ for different values of $\mu$. Right: $\omega_{ZO}(q)$ for different values of $\mu$. In both cases the coeficients of the analytic part of the dispersion are those obtained from the experimental fits in the main text.}\label{muDep}
\end{center}
\end{figure}

These self-energies can be then fed into Eq.~\ref{omega1} to determine the phonon energies. As an example, in Fig.~\ref{muDep} we have plotted $\omega_{LO}$ and $\omega_{ZO}$ for different values of $\mu$, with the coeficients of the analytic part as obtained from the fits to the experiment. This plots emphasize the dependence of the cusp position on the chemical potential, or equivalently, on $k_F$. 

\section{List of acronyms used}

\begin{itemize}
\item{EPC - Electron-phonon coupling}
\item{ZA - Out-of-plane acoustical}
\item{TA - Transverse acoustical}
\item{LA - Longitudinal acoustical}
\item{ZO - Out-of-plane optical}
\item{TO - Transverse optical}
\item{LO - Longitudinal optical}
\item{HREELS - High-resolution electron energy loss spectroscopy}
\item{ARPES - Angle-resolved photoemission spectroscopy}

\end{itemize}

\end{document}